\documentstyle[12pt]{article}
\begin{document}

\vspace{4cm}

\def\thefootnote{\fnsymbol{footnote}}
\begin{center}
{\large \bf
Metric on Quantum Spaces} \\

\ \\
\ \\

Andrzej Sitarz \footnote{Partially supported
by KBN grant 2P 302 168 4}
\footnote{E-mail: sitarz@if.uj.edu.pl} \\
Department of Field Theory \\
Institute of Physics \\
Jagiellonian University \\
Reymonta 4, 30-059 Krak\'ow, Poland
\end{center}

\vfill
\begin{abstract}

We introduce the analogue of the metric tensor in case
of $q$-deformed differential calculus. We analyse the
consequences of the existence of the metric, showing
that this enforces severe restrictions on the parameters
of the  theory. We discuss in detail the examples of the
Manin plane and the $q$-deformation  of $SU(2)$. Finally
we touch the topic of relations with the Connes' approach.

\end{abstract}

\vfill
\newpage

\section{Introduction}

Quantum groups and quantum spaces are an interesting non-trivial
generalization of Lie groups and manifolds \cite{qg}. The deformation
parameter $q$ allows to recover the latter in the continuous limit
$q \to 1$, which suggests that the noncommutativity of space could
possibly provide a regularization mechanism \cite{ma}. Therefore one
may expect that $q$-deformations could be an interesting basis for
physical theories, in particular for gravity and gauge theories
\cite{qgth}. The natural language for such studies is the $q$-deformed
differential calculus (see \cite{zum} for a review) constructed
within the framework of noncommutative differential geometry
\cite{con}. The construction, however is not unique, and even
after imposing bicovariance in the situation of quantum groups, in
general we are left with many choices of possible theories.

\def\a{{\cal A}}
The additional problem, which arises in the course of constructing
physical $q$-deformed theories is the question of metric. In the
classical situation, $q=1$, the metric is given by the metric
tensor. The latter could be equivalently defined as a bilinear
functional from $\Omega^1 \times \Omega^1$ to $\a$, where $\a$ is
the original commutative algebra, and $\Omega^1$ is the bimodule of
one-forms over $\a$. Now, we may generalize it to the case of
noncommutative geometry and define metric as a middle-linear
functional $\eta: \Omega^1 \times \Omega^1 \to \a$, i.e. $\eta$ is
linear with respect to addition in $\Omega^1$ and satisfies:
\begin{equation}
\eta( a \omega b, \rho c) \; = \; a \eta(\omega, b \rho) c,
\label{ml}
\end{equation}
for every $a,b,c \in \a$ and $\omega, \rho \in \Omega^1$.The middle
linearity naturally replaces the bilinearity condition in this
case, however, we shall see that it is far more restrictive. The
above definition proved to be suitable in our studies of discrete
geometries \cite{ja}.

We may also introduce the notion of hermitian metric, which could
be defined on the differential algebra with involution. We say
that metric $\eta$ is hermitian, if for all one-forms $u,v$ the
following identity holds:

\begin{equation}
\eta( u, v) \; = \; ( \eta(v^\star, u^\star) )^\star.
\label{her}
\end{equation}

In this paper we shall briefly discuss the consequences of the
introduction of the metric to the analysis of $q$-deformed
theories. We shall concentrate on two simple examples, leaving the
general case for future studies \cite{jedr}. Finally, we shall
discuss the relations between the above definition of the metric
and the approach of Connes \cite{con}.

\section{Metric on the Manin Plane}

Let us remind that the Manin plane is defined by replacing the
commutativity of the generators of $\c[x,y]$ by the relation:
\begin{equation}
 x y \; = \; q y x,
\label{mp1}
\end{equation}
where $q$ is a complex unitary number, $q\bar{q}=1$. We restrict
our considerations here to the algebra obtained as a quotient of
the free algebra generated by $y$ and $y$ by the ideal set by the
relation (\ref{mp1}).

Now, let us consider a $GL(2)_q$ invariant differential calculus
\cite{brzez}. It has one free parameter $s$ and its multiplication
rules are as follows:

\begin{eqnarray}
x dx &=& s dx x, \\
x dy &=& (s-1) dx y + q dy x, \\
y dx &=& s q^{-1} dx y, \\
y dy &=& s dy y.
\end{eqnarray}

The metric, due to the middle-linearity, is completely determined
by its values on the forms $dx$ and $dy$. If we call them
$\eta^{xx}$
for $\eta(dx,dx)$, and $\eta^{xy},\eta^{yx},\eta^{yy}$ for other
combinations respectively, we find the following
set of constraints:
\begin{eqnarray}
x \eta^{xx} & = & s^2 \eta^{xx} x, \label{r1} \\
y \eta^{xx} & = & s^2 q^{-2} \eta^{xx} y \\
x \eta^{xy} & = & s(s-1) \eta^{xx} y + sq \eta^{xy} x \\
y \eta^{xy} & = & s^2 q^{-1} \eta^{xy} y \\
x \eta^{yx} & = & s(s-1) q^{-1} \eta^{xx} y +  sq \eta^{yx} x \\
y \eta^{yx} & = & s^2 q^{-1} \eta^{yx} y \\
x \eta^{yy} & = & s(s-1) \eta^{xy} y + (s-1) q \eta^{yx} y + q^2
\eta^{yy} x \\
y \eta^{yy} & = & s^2 \eta^{yy} y, \label{r2}
\end{eqnarray}

Now, if we analyse them  we find restrictions on the
parameters $s$ and $q$. Since only monomials satisfy the
commutation
relations of the type (\ref{mp1}), we come to conclusion that in
the relation (\ref{r1}) $s^2$ must be equal to $q^n$ for some $n
\geq 0$.
Analogously, from (\ref{r2}) we see that $s^2$ must be $q^{-m}$ for
$m \geq 0$. Therefore, either  $n=m=0$ and hence $s^2=1$ or
$q^{n+m}=1$
and $s$ is any of the powers of $q$. In either case the possible
values of $s$
are restricted to a finite set. In particular, we have found that
out of infinitely many models of $q$-deformed, $GL(2)_q$ invariant
differential calculus on the Manin plane, only two of them admit
a metric for every value $q$.

\section{Metric on $SU(2)_q$}

The Hopf algebra of $SU(2)_q$ is generated by two elements (and
their conjugates), satisfying the following relations \cite{Woro}:
\begin{center}
\begin{tabular}{{c}@{\hspace{3cm}}{c}}
$a^\star a  + b^\star b = 1$ & $a a^\star + q^2 b^\star b = 1$
\end{tabular}

\begin{tabular}{{c}@{\hspace{1.5cm}}{c}@{\hspace{1.5cm}}{c}}
$b^\star b = b b^\star$  & $ab = q ba $ & $a b^\star = q b^\star a$
\end{tabular}
\end{center}
where $q \in [-1,1]$.

Following Woronowicz \cite{Woro} we introduce the bimodule of
one-forms, to be a free right-module generated by three elements
$c^0,c^1,c^2$, with the following rules of left multiplication
by the generator of $SU(2)_q$:\\[0.5cm]

\begin{tabular}{{l}@{\hspace{1.5cm}}{l}@{\hspace{1.5cm}}{l}}
$c^0 a =  q^{-1} a c^0$ & $c^1 a = q^{-2} a c^1$ & $c^2 a = q^{-1}
a
c^2$ \\
$c^0 b =  q^{-1} b c^0$ & $c^1 b = q^{-2} b c^1$ & $c^2 b = q^{-1}
b
c^2$ \\
$c^0 a^\star =  q a^\star c^0$ & $c^1 a^\star = q^{2} a^\star c^1$
&
$c^2 a^\star = q a^\star c^2$ \\
$c^0 b^\star =  q b^\star c^0$ & $c^1 b^\star = q^{2} b^\star c^1$
&
$c^2 b^\star = q b^\star c^2.$
\end{tabular}\\[0.5cm]

Additionally, the involution is extended to the bimodule of one
forms and we have:\\[0.5cm]

\begin{tabular}{{c}{c}{c}}
$(c^0)^\star = q c^0$ &   $(c^1)^\star = - c^1$ & $(c^2)^\star =
q^{-1} c^2.$
\end{tabular}\\[0.5cm]

Suppose now that we introduce the metric $\eta$, as proposed in the
first section. Since the module of one-forms is free, the metric
is completely determined by its values on the one-forms forming the
basis. Let us call $\eta(c^i,c^j)$ by $\eta^{ij}$, $i=0,1,2$.
Then, if we impose the condition of middle-linearity, we obtain
the following set of relations for the elements $\eta^{ij}$:

\begin{eqnarray}
a \eta^{ij} &=& q^{\phi(i,j)} \eta^{ij} a,  \\
a^\star \eta^{ij} &=& q^{-\phi(i,j)} \eta^{ij} a^\star,  \\
b \eta^{ij} &=& q^{\phi(i,j)} \eta^{ij} b, \label{rr1} \\
b^\star \eta^{ij} &=& q^{-\phi(i,j)} \eta^{ij} b^\star,
\label{rr2}
\end{eqnarray}

where $\phi(i,j)$ is defined as:

$$ \phi(i,j) = \cases{ 4 & if $i=j=1$ \cr
3 & if $i \not= j =1$ or $j \not= i =1$ \cr
2 & if $i \not= 1$ and $j \not=1$}. $$

One may easily verify that in the considered algebra such
constraints may be satisfied only if $q^2=1$. Indeed, from the
relations (\ref{rr1}) and (\ref{rr2}) we obtain that
$q^{2 \phi(i,j)}=1$. By taking all possible values of $i$ and $j$
we recover the above condition $q^2=1$. In the only non-trivial
case $q=-1$ we could have, for instance, the following metric:
\begin{equation}
\eta^{ij} \; = \; ab , \;\;\; \hbox{for} \;\; i \not= j =1 \;\;
\hbox{or} \;\; j \not= i =1,
\end{equation}
and the other components taken as constants. Of course, we could
scale each component by an arbitrary element of the center of the
algebra.

In the case of $SU(2)_q$ the existence of the metric is a
very strong requirement, which practically determines the value of
the
deformation parameter $q$ in the considered example of the
differential
calculus.

\section{Conclusions}

As we have shown in two previous sections, the existence of a
non-trivial metric is, in general, a very strong assumption.
We have demonstrated that the noncommutativeness of the original
algebra as well as of the differential calculus, enforce severe
restrictions on the possible metrics. They could not be satisfied
in general and lead to the constraints on the free parameters of
theory. Therefore some models of differential calculus seem to be
selected in a natural way by admitting the existence of the metric.
It should be therefore interesting to determine such relations for
other models, in particular for the general case of the bicovariant
differential calculus on quantum groups.

Having defined the metric one could also use the construction to
pursue the physical aspect of $q$-deformed theories. The natural
next step should be the introduction of linear connections and
$q$-deformed gravity, which is the topic of our current
investigation \cite{jedr}.

\section{Appendix}

In Connes approach, the basic object is a $K$-cycle, defined by
the algebra ${\cal A}$, its  representation $\pi$ on a Hilbert
space and the Dirac operator $D$. The differential algebra could be
derived from this construction by extending the definition of $\pi$
to the universal differential algebra $\Omega({\cal A})$:
\begin{equation}
\pi(a^0 da^1 \ldots da^n) \; = \; a^0 [D,a^1] \ldots [D,a^n],
\end{equation}
and by dividing $\Omega({\cal A})$ by the differential ideal
$\pi^{-1}(0) + d \pi^{-1}(0)$.

Now, introducing a Dixmier trace, we have the integration
on ${\cal A}$:
\begin{equation}
\int  a \; = \; \hbox{Tr} \, {\pi}(a),
\end{equation}
as well as a complex valued functional on the bimodule of one-
forms:
\begin{equation}
\langle u, v \rangle \; = \; \hbox{Tr}( \pi(u) \pi(v) ).
\label{ff}
\end{equation}

Let us turn back to the situation we were analysing. If we have a
hermitian metric $\eta$ and a positive trace (integration $\int$)
on the algebra ${\cal A}$ (which could be equivalent to the Dixmier
trace), we can recover the functional of the type (\ref{ff}) as
follows:
\begin{equation}
\langle u, v \rangle \; = \; \int \eta(u,v),
\end{equation}

Now, the interesting question is whether the existence of the
metric for a given differential calculus over ${\cal A}$ is
equivalent to the existence of the corresponding $K$-cycle over
${\cal A}$. If so, we could use the results of our studies of the
metric tensor in noncommutative geometry also in the broader
context. This should provide us with a link, which would enable to
extend the discrete geometry formalism of the Standard model
\cite{CL} to include also the gravitational component. Additionally,
we could then proceed with the introduction of $q$-deformed spinors,
attempting to deform physical models of fundamental matter fields.

\end{document}